\documentclass[manuscript,screen,nonacm]{acmart}

\acmConference[]{} 

\newcommand{\workshopname}{GenAICHI: CHI 2024 Workshop on Generative AI and HCI}
\newcommand{\licensedetails}{Licensed under a Creative Commons Attribution 4.0 International License (CC BY 4.0). Copyright remains with the author(s).}
\newcommand\extrafootertext[1]{
    \bgroup
    \renewcommand\thefootnote{\fnsymbol{footnote}}%
    \renewcommand\thempfootnote{\fnsymbol{mpfootnote}}%
    \footnotetext[0]{#1}%
    \egroup
}

\AtBeginDocument{ 
    \fancypagestyle{firstpagestyle}{
        \fancyhf{}
        \fancyfoot[L]{\sffamily\footnotesize \workshopname}%
        \fancyfoot[C]{\sffamily\footnotesize \thepage}
    }
    \fancyhf{}
    \fancyhead[L]{\sffamily\footnotesize\shorttitle}
    \fancyhead[R]{\sffamily\footnotesize\shortauthors}
    \fancyfoot[L]{\sffamily\footnotesize\workshopname}%
    \fancyfoot[C]{\sffamily\footnotesize\thepage}
    \extrafootertext{\licensedetails}
}

\usepackage{listings}
\usepackage{colortbl}
\usepackage{algorithm}
\usepackage{algpseudocode}

\newif\ifsubmit
\submitfalse
\ifsubmit
    \newcommand{\stefanie}[1]{}
    \newcommand{\mackenzie}[1]{}
    \newcommand{\faraz}[1]{}
    \newcommand{\ahmed}[1]{}
    \newcommand{\changes}[1]{}

\else
    \newcommand{\stefanie}[1]{{\leavevmode\color[rgb]{1.0, 0.0, 0.5}{#1}}}
    \newcommand{\mackenzie}[1]{{\leavevmode\color[rgb]{1.0, 0.6, 0.0}{#1}}}
    \newcommand{\faraz}[1]{{\leavevmode\color[rgb]{0.5, 0, 1.0}{#1}}}
    \newcommand{\changes}[1]{{\leavevmode\color[rgb]{0.0, 0.0, 0.0}{#1}}}
    \newcommand{\ahmed}[1]{{\leavevmode\color[rgb]{0, 0.7, 1.0}{#1}}}
     
\fi

\usepackage{soul}
\usepackage{booktabs}
\usepackage{multirow}
\usepackage{graphicx}
\usepackage{microtype}
\usepackage{hyperref}

\begin{document}




\title{Shaping Realities: Enhancing 3D Generative AI with Fabrication Constraints}


\author{Faraz Faruqi}
\email{ffaruqi@mit.edu}
\affiliation{%
  \institution{MIT CSAIL}
  \city{Cambridge}
  \state{MA}
  \country{USA}
}

\author{Yingtao Tian}
\email{alantian@google.com}
\affiliation{%
  \institution{Google DeepMind}
  \city{Tokyo}
  \country{Japan}
}

\author{Vrushank Phadnis}
\email{vrushank@google.com}
\affiliation{%
  \institution{Google Research}
  \city{Mountain View}
  \state{CA}
  \country{USA}
}

\author{Varun Jampani}
\email{varunjampani@gmail.com}
\affiliation{%
  \institution{Stability AI}
  \city{Cambridge}
  \state{MA}
  \country{USA}
}


\author{Stefanie Mueller}
\email{stefanie.mueller@mit.edu}
\affiliation{%
  \institution{MIT CSAIL}
  \city{Cambridge}
  \state{MA}
  \country{USA}
}

\renewcommand{\shortauthors}{Faruqi et al.}



\begin{abstract}
Generative AI tools are becoming more prevalent in 3D modeling, enabling users to manipulate or create new models with text or images as inputs. This makes it easier for users to rapidly customize and iterate on their 3D designs and explore new creative ideas. These methods focus on the aesthetic quality of the 3D models, refining them to look similar to the prompts provided by the user. However, when creating 3D models intended for fabrication, designers need to trade-off the aesthetic qualities of a 3D model with their intended physical properties. To be functional post-fabrication, 3D models have to satisfy structural constraints informed by physical principles. Currently, such requirements are not enforced by generative AI tools. This leads to the development of aesthetically appealing, but potentially non-functional 3D geometry, that would be hard to fabricate and use in the real world.
This workshop paper highlights the limitations of generative AI tools in translating digital creations into the physical world and proposes new augmentations to generative AI tools for creating physically viable 3D models. We advocate for the development of tools that manipulate or generate 3D models by considering not only the aesthetic appearance but also using physical properties as constraints. This exploration seeks to bridge the gap between digital creativity and real-world applicability, extending the creative potential of generative AI into the tangible domain.



 \end{abstract}

%

\ccsdesc[500]{Human-centered computing~Human computer interaction (HCI)}


\keywords{generative AI; digital fabrication.}


\begin{teaserfigure}
\centering
  \includegraphics[width=\textwidth]{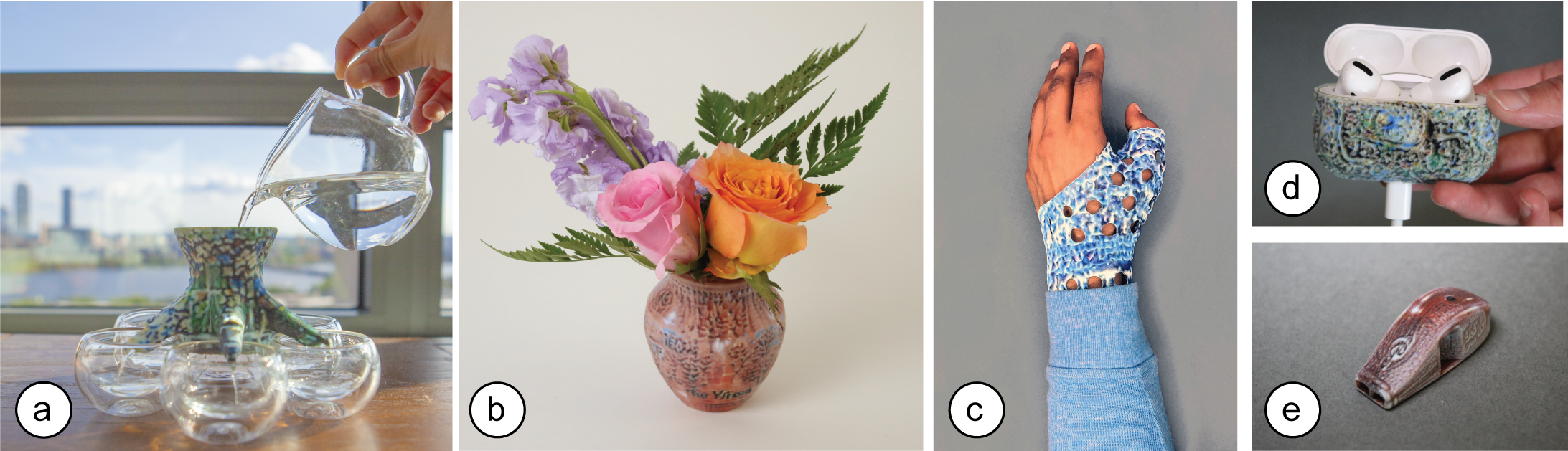}
  \vspace{-10pt}
  \caption{Using Generative AI for creating functional objects requires augmenting AI systems to preserve or encode functionality in 3D models. This will allow creators to generate personalized 3D models that not only have their desired aesthetic attributes but also their intended functionality. Here, we show some examples of open-sourced 3D models personalized with Generative AI while preserving their functionality and then 3D printed.  
  (a)~a drinks dispenser model styled as ``\textit{made of vintage mosaic glass tiles}''. (b)~A vase styled as ``\textit{a Terracotta vase}''. (c)~A personalized Thumb Splint styled like ``\textit{a blue knitted sweater}''. (d)~A personalized AirPods cover ``\textit{in the style of Moroccan Art}''.
  (e)~A functional whistle styled as ``\textit{A beautiful whistle made of mahogany wood}''.}
  \label{fig:teaser}
\end{teaserfigure}

\maketitle


\section{Introduction}

Generative AI is a rapidly expanding field and is transforming content creation processes. Tools leveraging generative AI enable users to easily and rapidly produce content; learning from varied data types like text, images, sounds, and 3D models. With an initial prompt in the form of a text description (for instance, “\textit{a vase with colorful flowers}”), or a seed image, the tool generates new content inspired by the training data. Innovative tools such as DALL-E~\cite{ramesh2021zero}, Stable-Diffusion~\cite{rombach2022high},  Muse~\cite{chang2023muse}, Magenta DDSP~\cite{engel2020ddsp}, Get3D~\cite{gao2022get3d}, Shap-E~\cite{jun2023shap_e}, Magic3D~\cite{lin2023magic3d} and Dream Gaussian~\cite{tang2023dreamgaussian} have facilitated the creation of new images, music, and high-resolution 3D models.

Novel methods~\cite{michel2022text2mesh, siddiqui2022texturify, jun2023shap_e, tang2023dreamgaussian} allow text and image-based manipulation of 3D models. This allows novice creators to explore open-source 3D models without necessitating learning complex 3D modeling techniques. However, these AI-driven co-creation tools are primarily oriented toward digital content creation. Their focus is on the tool's aesthetic and reconstruction capability; as demonstrated by their optimization method that maximizes apparent visual quality. Translation of these digital designs into tangible and physical forms via fabrication~\cite{baudisch2017personal,hudson2016understanding,yildirim2020digital, faruqi2021slicehub} remains an open problem. 

Designing 3D models for fabrication requires a trade-off between the aesthetic qualities of the design and its functional aspects. For example, designing a personalized thumb splint involves structural requirements such as ensuring rigidity for appropriate support for effective healing, or creating a movable lamp requires maintaining static equilibrium under various configurations. Expert designers optimize for both physical constraints and aesthetic goals while creating physically viable models. In digital fabrication, creators with little or no manufacturing experience often try to reuse previously shared designs on online platforms and send them for fabrication to a 3D printer~\cite{Kuznetsov_2010_expertamateur}. However, during customization, its important to ensure the new design has the functional properties they expect from the model. Conventional structural analysis tools conduct tests such as Finite element analysis (FEA) to reveal how a customized version of a model could fail. However setting up a model for structural analysis requires expertise in software tools, knowledge of structural analysis principles, and high computational power. They pose a barrier for non-expert creators who do not have specialized knowledge of the underlying principles or know ways to fix the failure cases from the simulation. 

Researchers have proposed a variety of ways to highlight violations of functional properties to non-expert users (e.g.,~\cite{hofmann2018greater, faruqi2023style2fab}). For instance, Style2Fab~\cite{faruqi2023style2fab} maintains physical functionality by using generative AI and a functionality-aware segmentation method to stylize the aesthetic parts of models without modifying the functional parts. However, functionality cannot always be localized and modifications made for aesthetic reasons may reduce the overall structural integrity of the model. This complex relationship between form and function in a 3D model requires a holistic analysis of the impact of a generative method on the physical properties of the model.   

This workshop paper aims to propose ideas for how generative AI systems can be advanced to support the creation of objects that are not only visually compelling but also functionally viable in the physical world. We discuss this idea in two different sections: \textit{Preserving Functionality} and \textit{Encoding Functionality}. \textit{Preserving Functionality} refers to how generative AI systems can be adapted to identify and preserve functional areas in a 3D model when manipulating aesthetic segments with text/image inputs. \textit{Encoding Functionality}, on the other hand, refers to encoding new functionality into a generated 3D model. Here, we present ideas on how learnings from the previous step can be used to guide generative AI methods to incorporate functionality while generating 3D models from scratch. Finally, we discuss the potential implications of our proposed methods on creators, maker space communities, and product design. 

\section{Preserving Functionality in 3D Models}
A key challenge for many makers is modifying or `stylizing' existing open-sourced designs shared in online repositories, such as Thingiverse~\cite{buehler2015sharing,alcock2016barriers}. These platforms allow users to browse and download from a large repository of existing models, but customization is often constrained to a limited set of pre-defined parameters. Moreover, these repositories often contain exported formats (OBJ/STL), which are hard to manipulate semantically as compared to CAD formats. Researchers have proposed tools such as Attribit~\cite{chaudhuri2013attribit} and Meshmixer~\cite{schmidt2010meshmixer} which allow limited manipulation of models based on a preprocessed set of semantic attributes, and models. However, generative AI tools offer an alternative. Methods like Text2Mesh~\cite{michel2022text2mesh} and Texturify~\cite{siddiqui2022texturify} allow manipulation of 3D models with text and image prompts as input and create stylized versions of the original model. However, these methods manipulate the entire 3D model, maximizing the apparent visual quality of the model to the desired style.

Beyond aesthetics, 3D printable models have functionality-motivated design parameters incorporated into the models. Editing such a design ad-hoc might compromise its functionality post-fabrication. Customization can be localized; however, that requires users to correctly identify and separate elements in a 3D model that have functionality associated with them. This would be a daunting task for makers without structural and mechanical knowledge of how different elements in geometry interact. Style2Fab~\cite{faruqi2023style2fab} presented an approach to augment this technique with automated identification of functional areas. This method segments a given model, identifies the functional areas, and then prevents any geometrical manipulations on these regions. This allows a user to manipulate the aesthetic regions of any model while preserving the functional regions. Some examples of objects created with this tool are shown in ~\autoref{fig:teaser}.



While this method is useful for separating functional regions, it leads to new open questions to answer: 

\begin{enumerate}

    \item \textbf{Identification of Functional Regions:} Functionality is highly context dependent. Similar geometrical segments could have different functionalities depending on the overall geometry and use cases of the user. Thus, to identify functional segments for a particular use case, any automatic classification algorithm has to also consider the context of the model in addition to the geometrical features.
    
    
    \item \textbf{Smart Manipulation of Functional Regions:} Style2Fab considers all functional segments out of bounds for any manipulation. Doing so allows the user to retain the functionality, but also carries risks of over-constraining the design space exploration. For example, the flat base of a vase can be manipulated, as far as the resulting surface results in a stable vase. Extending current tools to be physics-aware can be helpful for a ``smart'' stylizing, which allows creators to manipulate 3D models safely while preserving the intended functionality.
    
    \item \textbf{Relationship between Functional and Aesthetic Segments:} By separating functional and aesthetic segments, prior methods assume that any manipulation of aesthetic segments pose no impact on the functional segments of the model. In an ideal setting, such an approach creates usable models, as the original functional segments are theoretically preserved. But in real-world scenarios, this assumption might not hold --- In contrast, designers often co-edit the stylistic and structural aspects of a 3D model~\cite{yang}. In the case of a thumb splint, a user can create a wearable splint by (1) preserving the inner section that makes contact with the skin, and (2) manipulating the external surface. However, the external and internal sections of the model are connected, hence manipulating one of them can easily reduce the structural strength of the model, creating a significantly weaker model. 
    
    
    \item \textbf{Material Specific Constraints}: The stylizing and structural analysis of a model are bound by its material of construction. Yield strength, hardness, and corrosion resistance are among a few examples of material properties that have a direct bearing on selected stylizing approach. It remains an open problem to extend current tools to consider these properties while making geometric manipulations on the 3D model. 
\end{enumerate}
\section{Encoding Physical Functionality in Generative 3D Models}
A critical aspect in the evolution of 3D modeling, especially within the context of generative AI, is encoding physical functionality into the generated designs. As AI-based tools are increasingly used for creating 3D models, a significant challenge emerges: ensuring that these models are not just visually appealing but also functionally viable when fabricated. The process involves integrating complex physical principles and material properties into the generative algorithms, which goes beyond the traditional focus on aesthetic design. Generative AI systems~\cite{tsalicoglou2024textmesh, tang2023dreamgaussian} are typically adept at creating detailed and intricate designs. However, without the encoding of physical functionality, these designs may lack practical applicability, especially when subjected to specific loading conditions including static and dynamic forces, pressure, and other environmental factors. Addressing this challenge involves several key considerations:



\begin{enumerate}
    \item \textbf{Incorporating Material Properties:} Understanding and encoding the properties of different materials into AI models is crucial. Different materials have different physical properties like tensile strength, elasticity, and surface hardness which can significantly affect the functionality of the final printed object.

    \item \textbf{Simulation and Testing within the AI Process:} Integrating simulation tools into the generative process can help in predicting how a design will perform under various conditions. This could involve stress testing, thermal analysis, and other simulations to ensure that the generated model will function as intended.
    
    \item \textbf{Complex Geometrical Considerations:} The encoding process must also account for the geometrical complexities of the models. This includes ensuring that moving parts can operate without interference, that structures are stable and balanced, and that the overall design is optimized for the intended function.

    \item \textbf{User-Centric Customization:} The generative process should allow for customization based on user requirements, which could vary widely. For instance, a prosthetic limb generated by AI must be customizable to fit the unique anatomical features of its user while ensuring comfort and functionality.
    
    \item \textbf{Feedback Loops for Continuous Improvement:} Implementing feedback mechanisms within AI systems using physical testing can allow for the continuous refinement of models based on functional performance. This could involve real-world testing feedback being used to iteratively improve the design.
\end{enumerate}

The goal of encoding physical functionality in generative 3D models is to bridge the gap between aesthetically driven design and practical, real-world applicability. This approach not only enhances the usability of the generated models but also opens up new possibilities for innovation in various fields where functionality is at the core of the design process. 



\section{Conclusion}

This workshop paper highlights the research opportunities in extending 3D generative tools by integrating physical constraints into the process. Such systems will enable the creation of functional designs with generative AI, allowing creators to prototype creative ideas for objects they can fabricate. We describe two focus areas: \textit{Preserving Functionality}, to maintain functionality when manipulating 3D models, and \textit{Encoding Functionality}, to encode functionality along with aesthetics when creating 3D models from scratch. By developing generative AI tools that create objects with intended functionality and aesthetic attributes, this research would open new avenues for innovation in personal fabrication and product design.



\section{Acknowledgments}
We thank the MIT-Google Program for Computing Innovation for their generous support. Furthermore, we thank Douglas Eck and Yuanzhen Li from Google for their support to this research.     




\bibliographystyle{ACM-Reference-Format}
\bibliography{references}

\end{document}